\documentclass{article}

\usepackage{PRIMEarxiv}

\usepackage[utf8]{inputenc} 
\usepackage[T1]{fontenc}    
\usepackage{hyperref}       
\usepackage{url}            
\usepackage{booktabs}       
\usepackage{amsfonts}       
\usepackage{nicefrac}       
\usepackage{microtype}      
\usepackage{lipsum}
\usepackage{fancyhdr}       
\usepackage{graphicx}       
\graphicspath{{media/}}     

\usepackage{amsmath,amssymb}
\usepackage{bm}
\newcommand{\rr}{{\bm{r}}}
\newcommand{\xx}{{\bm{x}}}
\newcommand{\pp}{{\bm{p}}}

\newcommand{\me}{{\rm{e}}}
\newcommand{\md}{{\rm{d}}}
\newcommand{\mi}{{\rm{i}}}
\newcommand{\GG}{{\bf{G}}}

\title{A Green-function unification of weak interactions with environmental dressing
}

\author{
  Johannes Fiedler \\
  Department of Physics and Technology\\
  University of Bergen\\
  All\'egaten 55, 5007 Bergen, Norway\\
  \texttt{johannes.fiedler@uib.no} \\
}

\begin{document}
\maketitle

\begin{abstract}
Weak, non-covalent interactions such as electrostatics, induction, and dispersion are commonly treated as distinct physical mechanisms, often using separate models and approximations. Here, we present a unified quantum-electrodynamical framework in which all weak interactions between neutral or charged subsystems emerge from a single interaction Hamiltonian and are expressed entirely in terms of electromagnetic Green functions.

Starting from macroscopic QED in arbitrary linear environments, we derive compact master formulas that generate permanent electrostatic, mixed induction, and fully induced (dispersion) interactions on equal footing. Monopoles, dipoles, and quadrupoles are treated consistently within the same multipole expansion, leading to a common structural form in which all interaction channels are obtained from contractions of multipole moments and response functions with Green tensors. Environmental effects enter universally through dressed Green functions.

The standard non-retarded free-space limits, including Keesom, Debye, and London/van der Waals interactions, are recovered as a consistency check of the formalism. Beyond this validation, the formulation shows that once the general expressions are established, different physical scenarios, such as interfaces, confinement, or polarisable continua, can be treated by direct substitution of the corresponding Green function, without reformulating the interaction for each case. This elevates Green functions to a universal organising principle for weak interactions in complex environments.
\end{abstract}


\section{Introduction}
\label{sec:introduction}

Weak, non-covalent interactions play a central role across physics, chemistry, and materials science, governing phenomena ranging from molecular recognition and solvation to soft-matter assembly and nanoscale forces. Despite their common electromagnetic origin, these interactions are traditionally classified and treated as distinct mechanisms, including electrostatic (Keesom), induction (Debye), and dispersion (London or Casimir--Polder) forces. As a result, their theoretical descriptions often rely on different models, levels of approximation, and computational strategies.~\cite{Israelachvili2011,Review2023}

From a fundamental perspective, all weak interactions originate from the coupling between charges, currents, and the electromagnetic field. Quantum electrodynamics (QED), therefore, provides a natural framework in which these interactions should, in principle, arise from a single interaction Hamiltonian~\cite{Milonni1994}. In practice, however, unifying electrostatics, induction, and dispersion within a common and operationally useful formalism has remained challenging, particularly in the presence of complex environments.

From a microscopic perspective, weak interactions have long been unified within molecular quantum electrodynamics (QED), where electrostatic, induction, and dispersion forces arise from a common interaction Hamiltonian in free space~\cite{CraigThiru,Salam2010,Buckingham1967,Margenau1937}. In this framework, electrostatic and induction contributions emerge from instantaneous Coulomb interactions, while dispersion forces arise from
electrodynamic coupling to the quantised field.

While this unified picture is well established in a vacuum, its extension to structured and dispersive environments is less straightforward. In particular, the consistent, unified incorporation of geometry, boundary conditions, and material response remains a challenge when different interaction mechanisms are modelled separately.

A powerful route to treating electromagnetic interactions in structured or dissipative media is provided by macroscopic QED~\cite{Scheel2008,Buhmann12a}, in which material response and geometry are encoded in electromagnetic Green's tensors. Within this approach, dispersion forces at interfaces~\cite{PhysRevA.92.043819,C9CP03165K}, within cavities~\cite{10.1063/5.0037629}, or in absorbing media~\cite{PhysRevA.60.4094,10.1063/5.0106503} can be expressed in terms of dressed Green functions, thereby avoiding the need for mode decompositions or ad hoc boundary conditions. More generally, macroscopic QED has been widely applied to interaction and energy-transfer processes in structured environments, including recent studies of resonance energy transfer in complex media~\cite{Buhmann2023,Buhmann2024a,Buhmann2024b}.

At the same time, electrostatic interactions and classical induction effects are typically formulated separately, often obscuring their relation to fluctuation-induced forces~\cite{Hinzmann2024}. Multipole-based descriptions have also been employed in other areas of physics, for example, in the classification of electronic states and transport phenomena in condensed-matter systems~\cite{PhysRevB.102.144441}. These approaches, however, pursue a different objective, focusing on symmetry and band-structure properties rather than on interaction energies.

This separation motivates the present work, in which we formulate a unified description of weak interactions within macroscopic QED, based on the Green-function structure of the electromagnetic field.

The central idea of the present work is to identify the common structural form underlying all weak electromagnetic interactions. In this formulation, interaction energies are expressed as contractions of multipole moments and response tensors with electromagnetic Green functions, which encode geometry, boundary conditions, and material response. As a consequence, once the general expressions are established, different physical scenarios, such as interfaces, cavities, structured media, or polarisable continua, can be treated by direct substitution of the corresponding Green function, without reformulating the interaction for each case.

In chemistry and molecular physics, environmental effects are frequently modelled using polarisable continuum models (PCM), in which a molecule is embedded in a dielectric medium and interacts with induced surface charges~\cite{PCM2005}. While PCM provides an efficient description of solvation and screening effects, its connection to a fully unified electromagnetic description of weak interactions is not always explicit, and different interaction classes are often corrected using different prescriptions.

In this work, we extend the unified description of weak interactions known from molecular QED in free space to macroscopic QED in arbitrary environments. Starting from a single-interaction Hamiltonian, we derive general master formulas that generate all interaction channels via systematic multipole expansions and perturbation theory. Permanent moments, induced response, and mixed contributions emerge naturally once the permanent--fluctuation split is applied, without the need to impose separate physical assumptions for different interaction types.

A key feature of the formulation is that environmental effects enter exclusively through electromagnetic Green functions. This provides a direct connection between macroscopic QED and continuum approaches in chemical physics, such as PCM, in which environmental effects are introduced via effective response models. Within the present framework, such descriptions arise naturally from the Green-function formalism. The present work focuses on the conceptual unification and on the point-particle limit; finite-size and cavity-specific effects will be addressed in a separate study.

The structure of the paper is as follows. In Sec.~\ref{sec:unified-framework}, we develop the unified interaction framework and derive the general master formulas for permanent, mixed, and induced interactions. In Sec.~\ref{sec:reproduce}, we demonstrate the consistency of the approach by reproducing the standard non-retarded free-space limits. Section~\ref{sec:excess} summarises the results and outlines possible extensions.

\section{Unified QED interaction framework}
\label{sec:unified-framework}

\subsection{Gauge principle, canonical variables, and environment}
\label{sec:gauge-env}

We consider a nonrelativistic set of ``active'' charged particles (atoms, molecules, ions) with coordinates $\{\hat{\rr}_\alpha\}$, momenta $\{\hat{\pp}_\alpha\}$, masses $\{m_\alpha\}$ and charges $\{q_\alpha\}$, embedded in an arbitrary linear, causal, possibly dispersive and absorptive environment described at the macroscopic level by $\varepsilon(\xx,\omega)$ and $\mu(\xx,\omega)$. A convenient starting point is macroscopic QED, where the passive medium degrees of freedom are encoded via reservoir fields and the field operators are expressed through the classical dyadic Green tensor~\cite{Scheel2008,Buhmann12a,Review2023}. For gauge-transformation and truncation-consistent constructions, we follow the contained-unitary viewpoint developed in Refs.~\cite{Stokes2021,Stokes2022,Stokes2023,Taylor2022,Arwas2023,Gustin2023}.

Gauge freedom is implemented by the usual local transformation
\begin{align}
  \bm A(\xx,t) &\to \bm A(\xx,t) + \nabla\chi(\xx,t)\,,\\
  \phi(\xx,t) &\to \phi(\xx,t) - \dot{\chi}(\xx,t) \,,
\end{align}
and a canonical quantisation requires a (partial) gauge fixing. In the following, we adopt the Coulomb gauge, in which the vector potential is transverse, $\nabla \cdot \bm A_\perp = 0$, and the scalar potential accounts for the instantaneous Coulomb interaction. This choice is convenient for separating longitudinal and transverse contributions, while the final physical results are gauge independent in the full theory.

The minimal-coupling Hamiltonian (spin and higher magnetic terms suppressed) reads
\begin{align}
  \hat H
  = \sum_\alpha \frac{1}{2m_\alpha}\!\left(\hat{\pp}_\alpha - q_\alpha \hat{\bm A}(\hat{\rr}_\alpha)\right)^2
  + \hat V(\{\hat{\rr}_\alpha\})
  + \hat H_{\rm field+med}
  + \hat H_{\rm Coul}\,,
  \label{eq:H_minimal_final}
\end{align}
where $\hat H_{\rm field+med}$ is the medium-assisted field Hamiltonian (reservoir representation)~\cite{Fiedler2017}
and $\hat H_{\rm Coul}$ collects instantaneous longitudinal contributions consistent with Gauss' law~\cite{Barcellona2017}.

\textit{Dyadic Green tensor.}
The classical dyadic Green tensor $\GG(\rr,\rr',\omega)$ is defined as the fundamental solution of the vector Helmholtz equation
\begin{align}
\Bigl[\nabla\times \mu^{-1}(\rr,\omega)\,\nabla\times
      -\frac{\omega^2}{c^2}\,\varepsilon(\rr,\omega)\Bigr]\GG(\rr,\rr',\omega)
= \mathbf I\,\delta(\rr-\rr')\,.
\label{eq:dyadic-helmholtz}
\end{align}
In macroscopic QED, field operators can be expressed in terms of $\GG$ and bosonic noise operators. All environment dependence then enters exclusively through $\GG$ (and its static limit).

\subsection{Power--Zienau--Woolley transformation as a unitary boost}
\label{sec:PZW}

A key step towards a unified treatment of weak interactions is to move between equivalent Hamiltonian representations by a unitary transformation that redistributes interaction terms between matter and fields. This is achieved by the Power--Zienau--Woolley (PZW) transformation~\cite{PowerZienau1959,Woolley1971}, which provides an equivalent multipolar representation of the light--matter interaction Hamiltonian. In the long-wavelength regime, the PZW unitary transformation may be written (at dipole order) as
\begin{align}
  \hat U_{\rm PZW} = \exp\!\left[-\frac{\mi}{\hbar}\,\hat{\boldsymbol\mu}\cdot\hat{\bm A}_\perp\right]\,,
  \label{eq:U_PZW_dipole_final}
\end{align}
with $\hat{\boldsymbol\mu}=\sum_\alpha q_\alpha(\hat{\rr}_\alpha-\bm R)$. $\bm R$ denotes the chosen multipole expansion point (typically the centre-of-mass for a neutral subsystem); physical observables are origin independent in the full theory, while truncated multipole expansions require a consistent choice. More generally, one may define
\begin{align}
  \hat U = \exp\!\left[-\frac{\mi}{\hbar}\int \md^3x\, \hat{\bm P}(\xx)\cdot \hat{\bm A}_\perp(\xx)\right]\,,
  \label{eq:U_PZW_general_final}
\end{align}
where the auxiliary polarisation satisfies $\nabla\cdot \hat{\bm P}(\xx)=-\hat\rho(\xx)$, while $\hat{\bm P}_\perp$ fixes the gauge family. This transformation has been widely used in molecular quantum electrodynamics and in discussions of gauge representations of light--matter interaction Hamiltonians~\cite{CraigThiru,Salam2010,Andrews2018,Woolley2020}.

Exact (untruncated) theories obey
\begin{align}
  \hat H_{\rm multipolar} = \hat U\,\hat H_{\rm minimal}\,\hat U^\dagger\,,
  \label{eq:gauge_equivalence_final}
\end{align}
i.e. a unitary equivalence between minimal and multipolar forms.

\subsection{Projection-consistent transformations}
\label{sec:contained}

When restricting to a truncated matter and/or field subspace, gauge equivalence is lost unless the unitary is contained in the same subspace~\cite{Stokes2021,Stokes2022,Taylor2022,Arwas2023,Gustin2023}. Let $\hat P$ denote the projector onto the chosen truncated Hilbert space. A projection-consistent unitary is
\begin{align}
  \hat U_{\hat P} \equiv \exp\!\left[-\frac{\mi}{\hbar}\,\hat P\,\hat S\,\hat P\right]\,,
  \qquad
  \hat S := \int \md^3x\, \hat{\bm P}(\xx)\cdot \hat{\bm A}_\perp(\xx)\,,
  \label{eq:U_contained_final}
\end{align}
and gauge-consistent truncated Hamiltonians are then defined in the usual way (BCH-consistent within $\hat P$), ensuring $\hat H^{(\hat P)}_{\rm multipolar}=\hat U_{\hat P}\hat H^{(\hat P)}_{\rm minimal}\hat U_{\hat P}^\dagger$.

\subsection{Unified multipole expansion of the active sources}
\label{sec:multipoles}

We start from the exact active charge and current densities
\begin{align}
  \hat\rho(\xx) = \sum_\alpha q_\alpha\,\delta\!\bigl(\xx-\hat{\rr}_\alpha\bigr)\,,\qquad
  \hat{\bm J}(\xx) = \frac{1}{2}\sum_\alpha \frac{q_\alpha}{m_\alpha}
  \left[\hat{\bm p}_\alpha\,\delta\!\bigl(\xx-\hat{\rr}_\alpha\bigr)
       + \delta\!\bigl(\xx-\hat{\rr}_\alpha\bigr)\,\hat{\bm p}_\alpha\right]\,.
  \label{eq:rhoJ_def_final}
\end{align}
Expand about a reference point $\bm R$ (e.g. molecular centre), with $\hat{\bm s}_\alpha=\hat{\rr}_\alpha-\bm R$.
Using
\begin{align}
  \delta(\xx-\hat{\rr}_\alpha)
  = \delta(\xx-\bm R) - \hat s_{\alpha i}\,\partial_i \delta(\xx-\bm R)
    + \frac{1}{2}\hat s_{\alpha i}\hat s_{\alpha j}\,\partial_i\partial_j\delta(\xx-\bm R)+\dots \,,
\end{align}
one obtains the Cartesian multipole moments
\begin{align}
  \hat Q &:= \sum_\alpha q_\alpha\,,\\
  \hat d_i &:= \sum_\alpha q_\alpha\,\hat s_{\alpha i}\,,\\
  \hat Q_{ij} &:= \sum_\alpha q_\alpha\left(3\hat s_{\alpha i}\hat s_{\alpha j}-\delta_{ij}\hat{\mathbf s}_\alpha^2\right)\,, \qquad \hat Q_{ii}=0\,,
  \label{eq:multipoles_def_final}
\end{align}
where $\hat Q_{ij}$ is the traceless electric quadrupole tensor.

\subsection{Longitudinal (electrostatic) Green function $G_{\rm es}$ and monopole-inclusive multipole Hamiltonian}
\label{sec:Ges}

Monopoles couple to the longitudinal (electrostatic) sector. Eliminating the scalar potential via Gauss' law, the longitudinal interaction energy can be written as a bilinear functional of $\hat\rho$ with an electrostatic Green function $G_{\rm es}$
\begin{align}
  \hat H_{\rm int}^{({\rm L})}
  = \frac12 \int \md^3x \int \md^3x'\;
    \hat\rho(\xx)\, G_{\rm es}(\xx,\xx')\, \hat\rho(\xx')\,.
  \label{eq:H_L_rhoGesrho_final}
\end{align}
In an inhomogeneous dielectric background, $G_{\rm es}$ solves the Poisson problem
\begin{align}
  \nabla\cdot\!\Bigl[\varepsilon(\rr,\omega\!\to\!0)\,\nabla G_{\rm es}(\rr,\rr')\Bigr]
  = -\delta(\rr-\rr')\,,
  \label{eq:Poisson_Ges}
\end{align}
with boundary conditions encoding interfaces, cavities (real-cavity/PCM), etc. This scalar Green function is related to the static dyadic Green tensor by
\begin{align}
  \GG_{\rm stat}(\rr,\rr')
  \equiv \lim_{\omega \to 0}\frac{\omega^2}{c^2}\GG(\rr,\rr',\omega)
  = - \nabla \nabla' \,G_{\rm es}(\rr,\rr')\,.
  \label{eq:Gstat_vs_Ges}
\end{align}
i.e. the electrostatic dyadic Green tensor is obtained from the rescaled low-frequency limit of the macroscopic Green tensor and equals the double gradient of the scalar electrostatic propagator~\cite{Barcellona2017}.

\subsection{Multipole-expanded longitudinal Hamiltonian}
Using the multipole expansion of $\hat\rho$ around $\bm R$,
\begin{align}
  \hat\rho(\xx) = \hat Q\,\delta(\xx-\bm R) - \hat d_i\,\partial_i \delta(\xx-\bm R) + \frac{1}{6}\,\hat Q_{ij}\,\partial_i\partial_j \delta(\xx-\bm R) + \dots\,,
  \label{eq:rho_multipole_cart_final}
\end{align}
and inserting into \eqref{eq:H_L_rhoGesrho_final} yields the standard electrostatic multipole series, but now environment-dressed via $G_{\rm es}$. For two subsystems $A,B$ with expansion points $\bm R_A,\bm R_B$ one finds, up to quadrupole order,
\begin{align}
  \hat H^{({\rm L})}_{AB}
  =&\;\frac12\Bigl[
     \hat Q_A \hat Q_B\, G_{\rm es}(\bm R_A,\bm R_B)
     - \hat Q_A \hat d_{B,j}\,\partial_{B,j} G_{\rm es}(\bm R_A,\bm R_B)
     - \hat d_{A,i}\hat Q_B\,\partial_{A,i} G_{\rm es}(\bm R_A,\bm R_B)
  \nonumber\\
  &\qquad
     + \hat d_{A,i}\hat d_{B,j}\,\partial_{A,i}\partial_{B,j} G_{\rm es}(\bm R_A,\bm R_B)
     + \frac{1}{6}\hat Q_A \hat Q_{B,ij}\,\partial_{B,i}\partial_{B,j} G_{\rm es}(\bm R_A,\bm R_B)
  \nonumber\\
  &\qquad
     + \frac{1}{6}\hat Q_{A,ij} \hat Q_B\,\partial_{A,i}\partial_{A,j} G_{\rm es}(\bm R_A,\bm R_B)
     - \frac{1}{6}\hat d_{A,k}\hat Q_{B,ij}\,\partial_{A,k}\partial_{B,i}\partial_{B,j} G_{\rm es}(\bm R_A,\bm R_B)
  \nonumber\\
  &\qquad
     - \frac{1}{6}\hat Q_{A,ij}\hat d_{B,k}\,\partial_{A,i}\partial_{A,j}\partial_{B,k} G_{\rm es}(\bm R_A,\bm R_B)
     + \dots
  \Bigr]\, .
  \label{eq:H_L_AB_multipoles_final}
\end{align}
Here $\partial_{A,i}$ acts on the first argument and $\partial_{B,j}$ on the second. Equation \eqref{eq:H_L_AB_multipoles_final} is the monopole-inclusive generator for all permanent electrostatic interactions in arbitrary environments, and it is the natural insertion point for polarisable continuum models~\cite{PCM2005}, respectively cavity screening~\cite{Fiedler2017} through $G_{\rm es}$~\cite{Barcellona2017}.

\subsection{Transverse multipolar coupling and unified interaction Hamiltonian}
\label{sec:transverse}

The remaining (radiative) coupling is transverse and is naturally expressed in multipolar form. Up to electric quadrupole order,
\begin{align}
  \hat H_{\rm int}^{(\perp)} = - \hat d_i\, \hat E_{\perp,i}(\bm R) - \frac{1}{6}\,\hat Q_{ij}\,\partial_i \hat E_{\perp,j}(\bm R) + \hat H_{\rm mag} + \hat H_{\rm self} + \dots \,,
  \label{eq:H_int_transverse_final}
\end{align}
where $\hat H_{\rm mag}$ collects magnetic dipole and higher magnetic terms and $\hat H_{\rm self}$ denotes the multipolar self-energy/$\bm P^2$-type terms generated by the transformation, which become crucial in gauge-consistent truncations~\cite{Stokes2021,Stokes2022,Taylor2022}.

\subsection{Unified monopole--dipole--quadrupole interaction generator}
Combining longitudinal and transverse sectors,
\begin{align}
  \hat H_{\rm int} = \hat H_{\rm int}^{({\rm L})} + \hat H_{\rm int}^{(\perp)}\,,
  \label{eq:H_int_total_final}
\end{align}
and Eqs.~\eqref{eq:H_L_rhoGesrho_final}--\eqref{eq:H_int_transverse_final} provide a single starting point from which all weak, non-covalent interactions emerge by perturbation theory, with environment dependence entering through $G_{\rm es}(\rr,\rr')$ for static/PCM-dressed electrostatics and $\GG(\rr,\rr',\omega)$ for induced interactions.

\subsection{Permanent vs. induced: operator split and master formulas}
\label{sec:perm-induced}

For any multipole operator $\hat{\mathcal M}\in\{\hat Q,\hat d_i,\hat Q_{ij},\dots\}$ we define
\begin{align}
  \hat{\mathcal M} = \underbrace{\mathcal M^{\rm (p)}}_{\text{permanent}=\langle\hat{\mathcal M}\rangle}
  + \underbrace{\delta\hat{\mathcal M}}_{\text{fluctuations}}\,,
  \label{eq:perm-fluct-split}
\end{align}
where the expectation value is taken in the (electronic) reference state of each subsystem (ground state or thermal state). Permanent interactions arise from products of $\mathcal M^{\rm(p)}$; induced interactions are generated by correlators and response functions associated with $\delta\hat{\mathcal M}$.

\textit{Master formula I: permanent (electrostatic) multipole interactions.}
From \eqref{eq:H_L_AB_multipoles_final} we may write the permanent electrostatic interaction compactly as
\begin{align}
  U^{\rm (perm)}_{AB} = \frac12 \sum_{m,n\in\{0,1,2\}}\frac{(-1)^{m+n}}{m!\,n!}\,\mathcal M^{\rm(p)}_{A,i_1\dots i_m}\,\mathcal M^{\rm(p)}_{B,j_1\dots j_n}\,\partial_{A,i_1}\!\cdots\partial_{A,i_m}\,\partial_{B,j_1}\!\cdots\partial_{B,j_n}\,G_{\rm es}(\bm R_A,\bm R_B)\,,
  \label{eq:master-perm}
\end{align}
with the identifications (in our Cartesian convention)
\begin{align}
  \mathcal M^{\rm(p)}_{A} &= Q_A^{\rm(p)}, \\
  \mathcal M^{\rm(p)}_{A,i} &= d^{\rm(p)}_{A,i}, \\
  \mathcal M^{\rm(p)}_{A,ij} &= \tfrac{1}{3}Q^{\rm(p)}_{A,ij} \,,
\end{align}
and $m$ and $n$ denote the multipole orders of particles A and B, respectively. Equation \eqref{eq:master-perm} is the most direct PCM-ready entry point: any cavity/interface dressing enters via $G_{\rm es}$ and its derivatives, with no need to treat charges/dipoles separately.

\textit{Master formula II: induced interactions from mQED correlators.}
Induced contributions follow from perturbation theory in $\hat H_{\rm int}^{(\perp)}$ and are expressed through field correlators, which in macroscopic QED are determined by $\GG$. In imaginary-frequency representation, the transverse electric-field correlator obeys
\begin{align}
  \bigl\langle \hat E_{\perp,i}(\rr)\,\hat E_{\perp,j}(\rr')\bigr\rangle
  = \frac{\hbar}{\pi\varepsilon_0}\int\limits_0^\infty \md\xi\;\xi^2\, G_{ij}(\rr,\rr', \mi\xi)\,,
  \label{eq:EEcorr_imfreq}
\end{align}
and gradients thereof generate quadrupolar couplings.

To keep notation uniform across multipole orders, we introduce generalised (electric) susceptibilities
$\bm{\chi}^{(m)}(\omega)$ that map transverse field gradients to induced multipoles,
\begin{align}
  \delta \hat{\mathcal M}_{i_1\dots i_m}(\omega)
  = \sum_{n\ge 1}\chi^{(m;n)}_{i_1\dots i_m\,;\,k_1\dots k_n}(\omega)\,
    \partial_{k_1}\!\cdots\partial_{k_{n-1}} \hat E_{\perp,k_n}(\bm R,\omega)\,,
  \label{eq:generalized-suscept}
\end{align}
where $m=1$ corresponds to dipole response (usual polarisability) and $m=2$ to quadrupolar response, etc.

In this compact language, the induced energy between two subsystems can be organised as a sum over multipole channels:
\begin{align}
  \lefteqn{U^{\rm (ind)}_{AB}
  = -\frac{\hbar}{2\pi\varepsilon_0}\int\limits_0^\infty \md\xi\;}\nonumber\\
  &\times
  \sum_{m,n\ge 1}
  \mathcal C^{(m,n)}:\Bigl[
  \bm{\chi}_A^{(m)}(\mi\xi)\;\cdot\;
  \nabla^{m-1}\GG(\bm R_A,\bm R_B,\mi\xi)\;\cdot\;
  \bm{\chi}_B^{(n)}(\mi\xi)\;\cdot\;
  \nabla'^{\,n-1}\GG(\bm R_B,\bm R_A,\mi\xi)
  \Bigr]\,,
  \label{eq:master-induced}
\end{align}
where $\mathcal C^{(m,n)}$ is a purely combinatorial tensor encoding the precise Cartesian contractions, and $\nabla^{m-1}$ indicates that $(m-1)$ gradients act on the first argument (and similarly for $\nabla'$). The tensor $\mathcal C^{(m,n)}$ encodes the purely combinatorial Cartesian contractions and normalisation factors arising from the multipole expansion. The symbol ``:'' denotes full contraction over all multipole and field indices, yielding a scalar interaction energy. Equation~\eqref{eq:master-induced} generates the fully induced (fluctuation-induced/dispersion) interactions for all multipole combinations, with environment dependence entirely contained in $\GG$. The related explicit expressions for the induced interactions can be found in App.~\ref{app:induced}.

\textit{Mixed permanent--induced terms.}
Mixed interaction terms arise when a permanent multipole moment of one subsystem induces a response in the other. Using the permanent--fluctuation split $\hat{\mathcal M}=\mathcal M^{\rm(p)}+\delta\hat{\mathcal M}$ inside perturbation theory generated by $\hat H_{\rm int}$, these contributions can be written in a unified form as
\begin{align}
  U^{\rm (mix)}_{AB}
  &=
  -\frac{1}{\varepsilon_0}
  \sum_{m\ge 0}\sum_{n\ge 1}
  \mathcal C^{(m,n)}:
  \Bigl[
  \mathcal M^{\rm(p)}_{A}
  \cdot
  \nabla^{m-1}\GG_{\rm stat}(\bm R_A,\bm R_B)
  \cdot
  \bm{\chi}^{(n)}_B(0)
  \cdot
  \nabla'^{\,n-1}\GG_{\rm stat}(\bm R_B,\bm R_A)
  \Bigr]
  \nonumber\\
  &\qquad + (A\leftrightarrow B),
  \label{eq:master-mixed}
\end{align}
where $\GG_{\rm stat}=\lim_{\omega\to 0}(\omega^2/c^2)\GG(\omega) = -\nabla\nabla' G_{\rm es}$ is the electrostatic dyadic Green tensor. Equation~\eqref{eq:master-mixed} generates all classical induction energies (monopole--dipole, dipole--dipole, dipole--quadrupole, etc.) in arbitrary environments within a single Green-function framework. Appendix~\ref{app:mixed} summarises all explicit permanent--induced interactions based on Eq.~\eqref{eq:master-mixed}.

\subsection{Summary and classification of interaction channels}
\label{sec:summary}

Equations~\eqref{eq:master-perm}, \eqref{eq:master-mixed}, and \eqref{eq:master-induced} together with Table~\ref{tab:channels} provide a complete classification of weak, non-covalent interactions generated by the unified QED interaction Hamiltonian~\eqref{eq:H_int_total_final}.

All interactions fall naturally into three classes: (i) permanent--permanent electrostatic interactions governed by the scalar Green function $G_{\rm es}$, (ii) mixed permanent--induced (classical induction) interactions governed by the electrostatic dyadic $\GG_{\rm stat}=-\nabla\nabla'G_{\rm es}$, and (iii) induced--induced (dispersion) interactions governed by the full macroscopic dyadic Green tensor $\GG(\rr,\rr',\omega)$.

\begin{table}[h!]
\centering
\renewcommand{\arraystretch}{1.3}
\begin{tabular}{c c c c c}
\hline\hline
$m$ & $n$ & Channel & Green object & Physical interpretation \\
\hline
\multicolumn{5}{c}{\textbf{Permanent--permanent (electrostatics)}}\\
\hline
0 & 0 & $Q$--$Q$ & $G_{\rm es}$ &
Coulomb / PCM-screened electrostatics \\

0 & 1 & $Q$--$d$ & $\nabla' G_{\rm es}$ &
Charge--dipole interaction \\

1 & 1 & $d$--$d$ & $\nabla\nabla' G_{\rm es}$ &
Permanent dipole--dipole (Keesom)\\

0 & 2 & $Q$--$Q_{ij}$ & $\nabla'\nabla' G_{\rm es}$ &
Charge--quadrupole \\

1 & 2 & $d$--$Q_{ij}$ & $\nabla\nabla'\nabla' G_{\rm es}$ &
Dipole--quadrupole \\

2 & 2 & $Q_{ij}$--$Q_{kl}$ & $\nabla\nabla\nabla'\nabla' G_{\rm es}$ &
Quadrupole--quadrupole \\[4pt]

\hline
\multicolumn{5}{c}{\textbf{Mixed permanent--induced (classical induction)}}\\
\hline
0 & 1 & $Q^{\rm(p)}$--$\delta d$ & $\nabla G_{\rm es}\,\nabla' G_{\rm es}$ &
Ion--induced dipole (Debye) \\

1 & 1 & $d^{\rm(p)}$--$\delta d$ & $\nabla\nabla' G_{\rm es}\,\nabla' G_{\rm es}$ &
Dipole--induced dipole \\

0 & 2 & $Q^{\rm(p)}$--$\delta Q_{ij}$ & $\nabla\nabla G_{\rm es}\,\nabla' G_{\rm es}$ &
Charge--induced quadrupole \\

1 & 2 & $d^{\rm(p)}$--$\delta Q_{ij}$ & $\nabla\nabla\nabla' G_{\rm es}\,\nabla' G_{\rm es}$ &
Dipole--induced quadrupole \\

2 & 1 & $Q_{ij}^{\rm(p)}$--$\delta d$ & $\nabla\nabla\nabla' G_{\rm es}\,\nabla' G_{\rm es}$ &
Quadrupole--induced dipole \\

2 & 2 & $Q_{ij}^{\rm(p)}$--$\delta Q_{kl}$ & $\nabla\nabla\nabla' G_{\rm es}\,\nabla' G_{\rm es}$ &
Quadrupole--induced quadrupole \\[4pt]

\hline
\multicolumn{5}{c}{\textbf{Induced--induced (dispersion / Casimir--Polder)}}\\
\hline
1 & 1 & $\delta d$--$\delta d$ & $\mathbf G$ &
Dispersion / Casimir--Polder \\

1 & 2 & $\delta d$--$\delta Q_{ij}$ & $\nabla'\mathbf G$ &
Higher-order dispersion \\

2 & 2 & $\delta Q_{ij}$--$\delta Q_{kl}$ & $\nabla\mathbf G\nabla'$ &
Quadrupolar dispersion \\
\hline\hline
\end{tabular}
\caption{Complete overview of interaction channels generated by the unified multipole Hamiltonian.
$m$ and $n$ denote the multipole rank of subsystems $A$ and $B$.
Permanent electrostatic interactions are governed by the scalar Green function $G_{\rm es}$.
Mixed permanent--induced (classical induction) terms are governed by the electrostatic dyadic
$\mathbf G_{\rm stat}=-\nabla\nabla'G_{\rm es}$.
Fully induced (dispersion) interactions are governed by the macroscopic dyadic Green tensor
$\mathbf G(\omega)$.}
\label{tab:channels}
\end{table}

At this point, the unification of weak, non-covalent interactions is complete. Equations~\eqref{eq:master-perm}, \eqref{eq:master-mixed}, and \eqref{eq:master-induced} provide a compact set of master expressions that generate electrostatic, induction, and dispersion interactions between neutral or charged subsystems.

All interaction channels are expressed in terms of multipole moments, response functions, and electromagnetic Green functions, without introducing separate physical models for different interaction types. Differences between interaction classes arise solely from the multipole rank and from whether permanent moments or induced response are involved.

A key consequence of this formulation is that environmental effects enter exclusively through the Green function. Once the general expressions are established, different physical scenarios, such as interfaces, confinement, cavities, or polarisable continua, can be treated by directly substituting the corresponding Green function, without reformulating the interaction for each
case.

As an example of this generality, Sec.~\ref{sec:excess} introduces generalised excess response functions, which provide a consistent description of embedded particles and regularise the divergences inherent to macroscopic electrodynamics.

In the following section, we use the free-space Green tensor to show that the general expressions reproduce the standard non-retarded limits.

\section{Reproducing known interactions in free space}
\label{sec:reproduce}

In this section, we demonstrate that the unified Green-function formulation developed in Sec.~\ref{sec:unified-framework} reproduces the standard non-covalent interactions in free space. To avoid ambiguities related to the order of spatial differentiation and low-frequency limits of electromagnetic Green functions, all results are derived from the full retarded bulk dyadic Green tensor. The non-retarded and electrostatic limits are taken only at the final stage.

\subsection{Bulk dyadic Green tensor}
\label{sec:G0}

In homogeneous medium ($\varepsilon(\omega)$, $\mu=1$), the dyadic Green
tensor solving
\begin{align}
\Bigl[\nabla\times\nabla\times-\frac{\omega^2}{c^2}\varepsilon(\omega)\Bigr]
\mathbf G_0(\bm r,\bm r',\omega)
=
\mathbf I\,\delta(\bm r-\bm r')\,,
\end{align}
is given by
\begin{align}
  \mathbf G_0(\bm r,\bm r',\omega)
  =
  \left(
  \mathbf I + \frac{1}{k^2(\omega)}\nabla\nabla
  \right)
  \frac{\me^{\mi k(\omega)\rho}}{4\pi\rho}\,,
  \qquad
  \rho=\|\bm r-\bm r'\|\,,
  \label{eq:G0_full}
\end{align}
with $k^2(\omega)=\varepsilon(\omega)\omega^2/c^2$. This expression contains both transverse (radiative) and longitudinal (Coulomb) contributions in a gauge-consistent manner.

For later convenience, we introduce the rescaled Green tensor
\begin{align}
  \bm\Gamma(\bm r,\bm r',\omega)
  =
  \frac{\omega^2}{c^2}\,\mathbf G_0(\bm r,\bm r',\omega)\,,
  \label{eq:Gamma_def_ch3}
\end{align}
which directly generates electric-field correlators in macroscopic QED.

\subsection{Derivative structure and limiting procedures}
\label{sec:limits}

All interaction channels generated in Sec.~\ref{sec:summary} involve spatial derivatives of $\mathbf G_0$ or $\bm\Gamma$. Derivatives are always taken at a fixed finite $\omega$. Only after all contractions are performed are the physical limits taken.

\textit{Non-retarded limit.}
For $k(\omega)\rho\ll 1$ one expands
\begin{align}
  \frac{\me^{\mi k\rho}}{\rho}
  =
  \frac{1}{\rho}
  + \mi k
  - \frac{k^2\rho}{2}
  + \mathcal O(k^3\rho^2)\,,
\end{align}
which yields the non-retarded Green tensor
\begin{align}
  \bm\Gamma_{\rm nr}(\bm r,\bm r',\omega)
  =
  -\frac{1}{4\pi\varepsilon(\omega)\rho^3}
  \bigl(\mathbf I-3\,\bm e_\rho\otimes\bm e_\rho\bigr)\,,
  \qquad
  \bm e_\rho=\frac{\bm r-\bm r'}{\rho}\,.
  \label{eq:Gamma_nr_ch3}
\end{align}

\textit{Electrostatic limit.}
The electrostatic dyadic Green tensor is defined by
\begin{align}
  \mathbf G_{\rm stat}(\bm r,\bm r')
  =
  \lim_{\omega\to 0}\bm\Gamma(\bm r,\bm r',\omega)
  =
  -\nabla\nabla' G_{\rm es}(\bm r,\bm r')\,,
\end{align}
where the scalar electrostatic Green function in free space is
\begin{align}
  G_{\rm es}(\bm r,\bm r')
  =
  \frac{1}{4\pi\varepsilon(0)\rho}.
\end{align}

The required derivatives of $1/\rho$ are standard:
\begin{align}
  \partial_i\partial'_j\frac{1}{\rho}
  =
  -\frac{3 e_i e_j-\delta_{ij}}{\rho^3},
  \label{eq:d2_ch3}
\end{align}
with higher derivatives obtained by repeated differentiation. These identities generate all monopole, dipole, and quadrupole couplings appearing in Sec.~\ref{sec:summary}.

\subsection{Permanent--permanent dipoles: Keesom interaction}
\label{sec:keesom}

We first consider permanent multipole moments. Inserting $G_{\rm es}=1/(4\pi\varepsilon(0)\rho)$ into the permanent master formula \eqref{eq:master-perm} yields the familiar electrostatic interactions.

For two permanent dipoles $\bm d_A^{\rm(p)}$ and $\bm d_B^{\rm(p)}$, one finds
\begin{align}
  U_{\rm Keesom}(\rho)
  =
  \frac{1}{4\pi\varepsilon(0)}
  \frac{
  \bm d_A^{\rm(p)}\!\cdot\!\bm d_B^{\rm(p)}
  -3(\bm d_A^{\rm(p)}\!\cdot\!\bm e_\rho)
   (\bm d_B^{\rm(p)}\!\cdot\!\bm e_\rho)
  }{\rho^3}\,,
  \label{eq:Keesom}
\end{align}
which is the standard non-retarded dipole--dipole interaction. All other permanent channels (charge--dipole, quadrupole interactions) are obtained analogously from higher derivatives of $G_{\rm es}$.

\subsection{Permanent--induced dipoles: Debye interaction}
\label{sec:debye}

Mixed permanent--induced interactions follow from the master formula \eqref{eq:master-mixed}. In a bulk system, they reduce to classical induction energies.

As a representative example, consider a permanent dipole $\bm d_A^{\rm(p)}$ inducing a dipole moment in subsystem $B$. The electric field generated by $\bm d_A^{\rm(p)}$ is
\begin{align}
  \bm E_A(\rho)
  =
  \frac{1}{4\pi\varepsilon(0)\rho^3}
  \left[
  3\bm e_\rho(\bm e_\rho\!\cdot\!\bm d_A^{\rm(p)})-\bm d_A^{\rm(p)}
  \right]\,.
\end{align}
The induced energy is therefore
\begin{align}
  U_{\rm Debye}(\rho)
  =
  -\frac{1}{2}\,
  \bm E_A\cdot\boldsymbol{\alpha}_B(0)\cdot\bm E_A\,,
  \label{eq:Debye}
\end{align}
which coincides exactly with the $m=n=1$ mixed channel obtained from \eqref{eq:master-mixed} using $\mathbf G_{\rm stat}=-\nabla\nabla'G_{\rm es}$. Charge--induced dipole and higher multipole induction terms follow in the same manner.

\subsection{Induced--induced dipoles: London dispersion}
\label{sec:london}

Finally, we consider fluctuation-induced interactions. For the dipole--dipole channel ($m=n=1$), the induced master formula
\eqref{eq:master-induced} yields
\begin{align}
  U_{\rm vdW}(\rho)
  =
  -\frac{\hbar}{2\pi\varepsilon_0}
  \int\limits_0^\infty \md\xi\;
  \alpha_A(\mi\xi)\alpha_B(\mi\xi)\,
  \mathrm{Tr}\!\left[
  \bm\Gamma_{\rm nr}(\bm R_A,\bm R_B,\mi\xi)\cdot
  \bm\Gamma_{\rm nr}(\bm R_B,\bm R_A,\mi\xi)
  \right]\,.
\end{align}
Using \eqref{eq:Gamma_nr_ch3}, one obtains
\begin{align}
  U_{\rm vdW}(\rho)
  =
  -\frac{3\hbar}{\pi}
  \frac{1}{(4\pi\varepsilon_0)^2\,\rho^6}
  \int\limits_0^\infty \md\xi\;
  \frac{\alpha_A(\mi\xi)\alpha_B(\mi\xi)}{\varepsilon(\mi\xi)^2}\,,
  \label{eq:London}
\end{align}
which is the standard non-retarded London dispersion interaction. Higher-order dispersion channels (dipole--quadrupole, quadrupole--quadrupole) follow from the same expression by including the corresponding spatial derivatives of $\bm\Gamma_{\rm nr}$.

\subsection{Discussion}
\label{sec:ch3_discussion}

The examples above demonstrate that the unified Green-function formulation reproduces all standard non-covalent interactions in free space when evaluated in the appropriate limits. Electrostatic (Keesom), induction (Debye), and dispersion (London) interactions emerge as different contractions of the same Green tensors, rather than as distinct physical mechanisms.

This confirms the internal consistency of the framework and prepares the ground for the inclusion of environments and confinement effects via dressed Green functions. The formalism also naturally contains the retarded regime, in which the interaction reduces to the well-known Casimir--Polder potential with its characteristic $R^{-7}$ distance dependence in free space.

\section{Generalised excess multipole model (local-field renormalisation)}
\label{sec:excess}

A major advantage of the unified Green-function formulation is that environmental and cavity effects can, in suitable limits, be absorbed into effective (``excess'') source moments and response functions. In this section, we formulate a generalised excess-multipole model that extends the familiar excess polarisability concept to monopoles, dipoles, quadrupoles, and higher multipoles in a uniform manner.

\subsection{Onsager real-cavity factorisation and vertex dressing}
\label{sec:onsager_factor}

Consider a small spherical vacuum cavity (Onsager real cavity) embedded in a homogeneous bulk medium with dielectric function $\varepsilon(\omega)$~\cite{doi:10.1021/ja01299a050}. For sources located sufficiently close to the cavity centre, and for cavity radius $a$ much smaller than all other relevant length scales, the medium-assisted Green tensor inside the cavity factorises as~\cite{PhysRevA.75.042109,Fiedler2017}
\begin{align}
  \mathbf G_{\rm cav}(\rr,\rr',\omega)
  \simeq R(\omega)\,\mathbf G_{\rm bulk}(\rr,\rr',\omega)\,,
  \label{eq:G_factor}
\end{align}
with the local-field factor
\begin{align}
  L(\omega)=\frac{3\varepsilon(\omega)}{1+2\varepsilon(\omega)}\,,
  \qquad
  R(\omega)=L(\omega)^2\,.
  \label{eq:L_R_def}
\end{align}
Equation~\eqref{eq:G_factor} may be interpreted as a vertex renormalisation: each matter--field coupling inside the cavity acquires a factor $L(\omega)$, so that any propagator connecting two such vertices yields $R(\omega)=L(\omega)^2$.

In the unified formulation, all interaction energies are built from: (i) multipole moments (permanent sources), (ii) multipole susceptibilities (induced response), and (iii) Green tensors connecting the corresponding vertices. Therefore, whenever \eqref{eq:G_factor} holds, cavity effects can be absorbed into effective (excess) multipoles and susceptibilities while evaluating the remaining geometry using bulk Green tensors.

\subsection{Definition of excess multipoles and excess susceptibilities}
\label{sec:excess_defs}

Let $\mathcal M^{\rm(p)}_{i_1\dots i_m}$ denote a permanent electric multipole of rank $m$ (with $m=0$ monopole charge, $m=1$ dipole, $m=2$ traceless quadrupole, etc.). We define the corresponding \emph{excess multipole} by a vertex dressing factor evaluated in the relevant (typically static) limit:
\begin{align}
  \mathcal M^{\ast\,{\rm(p)}}_{i_1\dots i_m}
  =
  L(0)\,\mathcal M^{\rm(p)}_{i_1\dots i_m},
  \qquad (m=0,1,2,\dots)\,.
  \label{eq:excess_perm_multipoles}
\end{align}
This definition ensures that any permanent--permanent interaction energy, which is bilinear in the permanent moments and linear in the static Green object, automatically acquires the correct factor $R(0)=L(0)^2$.

For induced responses, we employ the multipole susceptibilities introduced in Sec.~\ref{sec:perm-induced}. In particular, for the electric dipole response $\bm\chi^{(1)}\equiv\bm\alpha(\omega)$ one recovers the standard excess polarisability concept,
\begin{align}
  \bm\alpha^\ast(\omega)=R(\omega)\,\bm\alpha(\omega)\,,
  \label{eq:alpha_excess}
\end{align}
leading to the well-known Onsager real-cavity model. More generally, for any electric multipole susceptibility of rank $m$, we define
\begin{align}
  \bm\chi^{(m)\ast}(\omega)
  =
  R(\omega)\,\bm\chi^{(m)}(\omega),
  \qquad (m=1,2,\dots)\,,
  \label{eq:excess_suscept}
\end{align}
i.e. susceptibilities acquire the full two-vertex factor $R(\omega)=L(\omega)^2$. This is the natural generalisation of \eqref{eq:alpha_excess} to quadrupolar and higher-order response.

\textit{Interpretation.}
Equations~\eqref{eq:excess_perm_multipoles}--\eqref{eq:excess_suscept} implement a
clean separation: bulk propagation is handled by $\mathbf G_{\rm bulk}$ (or $G_{{\rm es,bulk}}$), while cavity/local-field physics is absorbed into renormalised (excess) source and response quantities.

\subsection{Scope and limitations}
\label{sec:excess_scope}

The factorisation \eqref{eq:G_factor} is controlled by the small-cavity limit and the assumption that the relevant source points remain close to the cavity centre. Outside this regime, the cavity modifies not only the amplitude but also the tensorial structure and spatial dependence of the Green function, and one must revert to the full cavity Green tensor. Nevertheless, within its domain of validity, the generalised excess-multipole model provides a powerful, essentially algebraic route to incorporate cavity and local-field effects into all interaction channels derived in Sec.~\ref{sec:summary}.

\section{Conclusions}
\label{sec:conclusions}

We have presented a unified Green-function formulation of weak, non-covalent interactions in which electrostatic, induction, and dispersion forces emerge from a single interaction Hamiltonian. By working within macroscopic QED and employing a systematic multipole expansion, we derived compact master expressions, Eqs.~\eqref{eq:master-perm}--\eqref{eq:master-induced}, which generate all interaction channels that treat monopoles, dipoles, and quadrupoles on equal footing and apply to both neutral and charged subsystems.

A central result is that all interaction classes are governed by electromagnetic Green functions.  Permanent electrostatic interactions are governed by the scalar electrostatic Green function, mixed permanent--induced interactions by its dyadic counterpart, and fully induced (dispersion) interactions by the frequency-dependent dyadic Green tensor. This establishes a common structural form in which geometry, material response, and environmental effects enter exclusively through the choice of the Green function. As a consequence, once the general expressions are established, different physical scenarios, including interfaces, confinement, cavities, or polarisable continua, can be treated by direct substitution of the corresponding Green function, without reformulating the interaction for each case. This provides a transparent link between QED-based approaches and continuum descriptions commonly used in chemical physics.

The reproduction of the standard non-retarded free-space limits (Keesom, Debye, and London interactions) serves as a consistency check of the formalism and demonstrates that the traditional classification of weak interactions is not fundamental, but rather emerges from different contractions of the same underlying Green tensors.

The present work establishes a conceptual foundation for treating weak interactions in complex environments within a single, transparent formalism. As an example of the applicability of the framework, Sec.~\ref{sec:excess} introduces generalised excess response functions, which account for the embedding of particles in a medium and provide a consistent regularisation of short-distance divergences in macroscopic electrodynamics. Extensions to structured media, interfaces, and continuum solvation models follow naturally through dressed Green functions. Finite-size and cavity-specific effects, which require additional geometric input, will be addressed in a separate publication.

%
%
\section*{Acknowledgments}
The author would like to thank Alexander Friedrich and Enno Giese for valuable discussions and critical insights that contributed to the development of this work.

\section*{Data availability}
No new data were generated in this work. Any data required to reproduce the results are available from the author upon reasonable request.

\bibliographystyle{unsrt}
\bibliography{bibi.bib}

\appendix

\appendix
\section{Explicit multipole expansion of the induced interaction}
\label{app:induced}

This appendix provides the explicit evaluation of the master formula \eqref{eq:master-induced} up to electric quadrupole order ($m,n\le 2$), using traceless Cartesian quadrupole tensors throughout.

\subsection{General starting point}

The induced interaction energy between subsystems $A$ and $B$ is given in \eqref{eq:master-induced}, where $\GG$ is the macroscopic dyadic Green tensor, $\bm{\chi}^{(1)}\equiv\boldsymbol{\alpha}$ is the electric dipole polarisability, and $\bm{\chi}^{(2)}\equiv\boldsymbol{\beta}$ the electric quadrupole polarisability. The symbol ``$:$'' denotes full Cartesian contraction over all multipole and field indices.

\subsection{Dipole--dipole channel ($m=n=1$)}

The leading induced interaction is the familiar dipole--dipole (Casimir--Polder / van der Waals) contribution
\begin{align}
  U^{(1,1)}_{AB}
  =
  -\frac{\hbar}{2\pi\varepsilon_0}
  \int\limits_0^\infty \md\xi\;
  \alpha^{A}_{ik}(\mi\xi)\,
  G_{k\ell}(\bm R_A,\bm R_B,\mi\xi)\,
  \alpha^{B}_{j\ell}(\mi\xi)\,
  G_{\ell k}(\bm R_B,\bm R_A,\mi\xi)\,.
  \label{eq:U11}
\end{align}
For isotropic particles, $\alpha_{ik}=\alpha\,\delta_{ik}$, this reduces to the standard trace formula $\propto \mathrm{Tr}\bigl[\GG\cdot\GG\bigr]$.

\subsection{Dipole--quadrupole channel ($m=1,n=2$)}

The dipole--quadrupole contribution reads
\begin{align}
  U^{(1,2)}_{AB}
  =
  -\frac{\hbar}{2\pi\varepsilon_0}
  \int\limits_0^\infty \md\xi\;
  \frac{1}{2}
  \alpha^{A}_{ik}(\mi\xi)\,
  G_{k\ell}(\bm R_A,\bm R_B,\mi\xi)\,
  \beta^{B}_{j_1 j_2;\ell}(\mi\xi)\,
  \partial'_{j_2} G_{\ell k}(\bm R_B,\bm R_A,\mi\xi)\,,
  \label{eq:U12}
\end{align}
where $\beta_{j_1 j_2;\ell}$ is symmetric and traceless in $(j_1 j_2)$. The factor $1/2$ originates from the multipole expansion and avoids double counting of symmetric gradient contributions. The reverse channel $U^{(2,1)}_{AB}$ is obtained by exchanging
$A\leftrightarrow B$ and primed/unprimed derivatives.

\subsection{Quadrupole--quadrupole channel ($m=n=2$)}

The quadrupole--quadrupole induced interaction is
\begin{align}
  U^{(2,2)}_{AB}
  =
  -\frac{\hbar}{2\pi\varepsilon_0}
  \int\limits_0^\infty \md\xi\;
  \frac{1}{4}
  \beta^{A}_{i_1 i_2;k}(\mi\xi)\,
  \partial_{i_2} G_{k\ell}(\bm R_A,\bm R_B,\mi\xi)\,
  \beta^{B}_{j_1 j_2;\ell}(\mi\xi)\,
  \partial'_{j_2} G_{\ell k}(\bm R_B,\bm R_A,\mi\xi)\,.
  \label{eq:U22}
\end{align}
The prefactor $1/4$ arises from the product of the $1/2$ factors associated with each traceless quadrupole expansion.

\subsection{Compact representation}

Collecting all contributions up to quadrupole order,
\begin{align}
  U^{\rm (ind)}_{AB}
  =
  U^{(1,1)}_{AB}
  + U^{(1,2)}_{AB}
  + U^{(2,1)}_{AB}
  + U^{(2,2)}_{AB}
  + \mathcal O(Q^3)\,.
  \label{eq:Uind_total}
\end{align}
All environment and cavity effects enter exclusively through the dyadic Green tensor $\GG(\rr,\rr',\mi\xi)$ and its spatial derivatives.


\section{Explicit multipole expansion of mixed permanent--induced interactions}
\label{app:mixed}

This appendix provides the explicit evaluation of the mixed (permanent--induced) interaction energy generated by the unified Hamiltonian \eqref{eq:H_int_total_final}. Throughout, we employ traceless Cartesian quadrupole tensors and the electrostatic dyadic Green function
\begin{align}
  \GG_{\rm stat}(\rr,\rr') = \lim_{\omega\to 0}\frac{\omega^2}{c^2}\GG(\rr,\rr',\omega) = -\nabla\nabla' G_{\rm es}(\rr,\rr')\,.
\end{align}

The mixed interaction energy is obtained by combining the permanent multipole moments of one subsystem with the linear response of the other, Eq.~\eqref{eq:master-mixed} with the dipole polarisability $\bm{\chi}^{(1)}\equiv\boldsymbol{\alpha}$ and the quadrupole polarisability $\bm{\chi}^{(2)}\equiv\boldsymbol{\beta}$.

\subsection{Charge--induced dipole ($m=0,n=1$)}

The interaction between a permanent monopole of subsystem $A$ and an induced dipole in $B$ is
\begin{align}
  U^{(0,1)}_{AB}
  =
  -\frac{1}{\varepsilon_0}
  Q_A^{\rm(p)}\,
  \partial_i G_{\rm es}(\bm R_A,\bm R_B)\,
  \alpha^{B}_{ij}(0)\,
  \partial'_j G_{\rm es}(\bm R_B,\bm R_A)\,,
  \label{eq:U01}
\end{align}
which corresponds to the familiar ion--induced dipole interaction, here fully dressed by the environment via $G_{\rm es}$.

\subsection{Permanent dipole--induced dipole ($m=1,n=1$)}

The mixed dipole--dipole contribution reads
\begin{align}
  U^{(1,1)}_{AB}
  =
  -\frac{1}{\varepsilon_0}
  d^{\rm(p)}_{A,i}\,
  \partial_i\partial'_k G_{\rm es}(\bm R_A,\bm R_B)\,
  \alpha^{B}_{k\ell}(0)\,
  \partial'_\ell G_{\rm es}(\bm R_B,\bm R_A)\,,
  \label{eq:U11_mix}
\end{align}
together with the symmetric counterpart obtained by exchanging $A\leftrightarrow B$. Equation~\eqref{eq:U11_mix} reproduces the classical induction energy between
a permanent dipole and a polarisable object in an arbitrary environment.

\subsection{Charge--induced quadrupole ($m=0,n=2$)}

A permanent monopole inducing a quadrupole moment gives
\begin{align}
  U^{(0,2)}_{AB}
  =
  -\frac{1}{\varepsilon_0}
  \frac{1}{2}
  Q_A^{\rm(p)}\,
  \partial_i\partial_j G_{\rm es}(\bm R_A,\bm R_B)\,
  \beta^{B}_{ij;k}(0)\,
  \partial'_k G_{\rm es}(\bm R_B,\bm R_A)\,,
  \label{eq:U02}
\end{align}
where $\beta_{ij;k}$ is symmetric and traceless in $(i,j)$. Such terms are usually neglected but arise naturally in the unified framework.

\subsection{Permanent dipole--induced quadrupole ($m=1,n=2$)}

The dipole--quadrupole mixed interaction is
\begin{align}
  U^{(1,2)}_{AB}
  =
  -\frac{1}{\varepsilon_0}
  \frac{1}{2}
  d^{\rm(p)}_{A,i}\,
  \partial_i\partial_j\partial'_k G_{\rm es}(\bm R_A,\bm R_B)\,
  \beta^{B}_{jk;\ell}(0)\,
  \partial'_\ell G_{\rm es}(\bm R_B,\bm R_A)\,,
  \label{eq:U12_mix}
\end{align}
plus the exchanged contribution $A\leftrightarrow B$.

\subsection{Permanent quadrupole--induced dipole ($m=2,n=1$)}

Conversely, a permanent quadrupole inducing a dipole yields
\begin{align}
  U^{(2,1)}_{AB}
  =
  -\frac{1}{\varepsilon_0}
  \frac{1}{2}
  Q^{\rm(p)}_{A,ij}\,
  \partial_i\partial_j\partial'_k G_{\rm es}(\bm R_A,\bm R_B)\,
  \alpha^{B}_{k\ell}(0)\,
  \partial'_\ell G_{\rm es}(\bm R_B,\bm R_A)\,.
  \label{eq:U21_mix}
\end{align}

\subsection{Quadrupole--induced quadrupole ($m=2,n=2$)}

The highest-order mixed term retained here is
\begin{align}
  U^{(2,2)}_{AB}
  =
  -\frac{1}{\varepsilon_0}
  \frac{1}{4}
  Q^{\rm(p)}_{A,ij}\,
  \partial_i\partial_j\partial'_k G_{\rm es}(\bm R_A,\bm R_B)\,
  \beta^{B}_{k\ell;m}(0)\,
  \partial'_m G_{\rm es}(\bm R_B,\bm R_A)\,,
  \label{eq:U22_mix}
\end{align}
together with the symmetric term $A\leftrightarrow B$. The prefactor $1/4$ arises from the product of the $1/2$ factors associated with the traceless quadrupole expansion.

\subsection{Summary}

Collecting all mixed contributions up to quadrupole order,
\begin{align}
  U^{\rm (mix)}_{AB}
  =
  \sum_{\substack{m=0,1,2\\ n=1,2}}
  U^{(m,n)}_{AB}
  + (A\leftrightarrow B)
  + \mathcal O(Q^3),
\end{align}
where all interaction channels are governed by the same electrostatic Green function $G_{\rm es}$ and its derivatives. Environmental, cavity, and PCM effects, therefore, enter the mixed (permanent--induced) interactions in exactly the same manner as for the purely permanent electrostatic terms.

\end{document}